\documentclass[prl,twocolumn,preprintnumbers,amsmath,amssymb]{revtex4-1}
\usepackage{graphicx}

\begin{document}
\title{ Parity oscillations and photon correlation functions in the $ Z_2/U(1) $ Dicke model at a finite number of atoms or qubits }
\author{  Yu Yi-Xiang$^{1,2,3}$, Jinwu Ye $^{1,2,4}$ and CunLin Zhang $^{2}$ }
\affiliation{ 
   $^{1}$  Department of Physics and Astronomy, Mississippi State  University, P. O. Box 5167, Mississippi State, MS, 39762   \\ 
   $^{2}$  Key Laboratory of Terahertz Optoelectronics, Ministry of Education, Department of Physics, Capital Normal University, Beijing 100048, China  \\
   $^{3}$  School of Instrument Science and Opto-electronics Engineering, Institute of Optics and Electronics, BeiHang University, Beijing 100191, China  \\
   $^{4}$  Kavli Institute of Theoretical Physics, University of California, Santa Barbara, Santa Barbara, CA 93106   }
\date{\today }


\begin{abstract}
  In this work, by using the strong coupling expansion
  and exact diagonization (ED), we study the $ Z_2/U(1) $ Dicke model  with independent rotating wave (RW) coupling $ g $
  and  counter-rotating wave (CRW) coupling $ g^{\prime} $  at a finite $ N $.
  This model includes the  four standard quantum optics model: Rabi, Dicke, Jaynes-Cummings ( JC ) and Tavis-Cummings (TC)  model
  as its various special limits.
  We show that in the super-radiant phase, the system's energy levels are grouped into doublets  with even and odd parity.
  Any anisotropy $ \beta=g/g^{\prime} \neq 1 $
  leads to the oscillation of parities in both the ground and excited doublets as the atom-photon coupling strength increases.
  The oscillations will be pushed to the infinite coupling strength in the isotropic $ Z_2 $ limit $ \beta=1 $.
  We find nearly perfect agreements between the strong coupling expansion and the ED in the super-radiant regime when $ \beta $ is not too small.
  We also compute the photon correlation functions, squeezing spectrum, number correlation functions which can be measured by various
  standard optical techniques.
\end{abstract}

\maketitle

  {\sl Introduction:}
  There are several well known quantum optics  models to study atom-photon interactions\cite{walls,scully}.
  In the Rabi model\cite{rabi}, a single mode photon interacts with a two level atom with equal rotating wave
  (RW) and counter rotating wave (CRW) strength.  When the coupling strength is well below the transition frequency,
  the CRW term is effectively much smaller than that of RW, so  it was dropped in the Jaynes-Cummings ( JC ) model \cite{jc}.
  Then the Rabi and JC model were extended to an assembly of $ N $ two level atoms to the Dicke model \cite{dicke}
  and the Tavis-Cummings (TC) model \cite{tc} respectively.
  Despite their relative simple forms and many previous theoretical works \cite{dicke1,popov,chaos,rabisol,qcphoton,infinite},
  their solutions at a finite $ N $, especially inside the superradiant regime, remain unknown.
  Here, we address this outstanding problem.
  It is convenient to classify the four well known quantum optics models from a simple symmetry point of view:
  the TC and Dicke model as the $ U(1) $ and $ Z_2 $ Dicke model  \cite{berryphase,gold,comment} respectively,
  while JC and Rabi model are just as the $ N=1 $ version of the two.

  Due to recent tremendous advances in technologies, ultra-strong couplings in cavity QED systems were achieved in at least
  two experimental systems (1) a BEC atoms
  inside an ultrahigh-finesse optical cavity \cite{qedbec1,qedbec2,orbitalt,orbital,switch}  and
  (2) superconducting qubits inside a microwave circuit cavity \cite{qubitweak,qubitstrong,ultra1,ultra2,dots}.
  In general, in such a ultra-strong coupling regime, the system is described well by
  Eq.\ref{u1z2z2} dubbed as the $ U(1)/Z_2 $ Dicke model \cite{gold,gprime1,expggprime}
  which includes the  four standard quantum optics model as its various special limits.
  Here, we study  the $ U(1)/Z_2 $ Dicke model Eq.\ref{u1z2z2} at any finite $ N $ and any ratio between the RW and
  the CRW term $ 0 \leq g^{\prime}/g =\beta \leq 1 $
  by the strong coupling expansion \cite{strongc} and exact diagonization (ED) \cite{chaos,gold,china}.
  We show that  in the super-radiant phase, the system's energy levels are grouped into doublets each of which
  consists of two Schrodinger Cat states with even and odd parity.
  Any anisotropy $ \beta \neq 1 $  leads to the oscillation of parities in the ground and excited doublet states in superradiant phase as the $ g $ increases.
  In the $ Z_2 $ limit $ \beta=1 $, all the oscillations are pushed to $ g =\infty $.
  We find nearly perfect agreements between the strong coupling expansion and the ED
  in the superradiant regime when $ \beta $ is not too  small.
  We  compute the photon correlation functions, squeezing spectrum and number correlation functions which can be
  detected by fluorescence spectrum, phase sensitive homodyne detection and Hanbury-Brown-Twiss (HBT)
  type of experiments respectively \cite{walls,scully,exciton}. Experimental realizations are discussed.
  New perspectives are outlined.


{\sl Strong coupling expansion--- }
  In the strong coupling limit, it is more convenient to
  rewrite the $ U(1)/Z_2 $ Dicke model  \cite{gold} in its  dual $ Z_2/U(1) $ presentation:
\begin{eqnarray}
  H_{Z_2/U(1)} &= & \omega_a a^{\dagger} a + \omega_b J_z + \frac{ g(1+\beta) }{ \sqrt{N} } ( a^{\dagger}+ a ) J_x  \nonumber  \\
 & - & \frac{ g(1-\beta) }{ \sqrt{N} } ( a^{\dagger} - a )i J_y
\label{u1z2z2}
\end{eqnarray}
   where  $ \omega_a, \omega_b $ are the cavity photon frequency and the energy difference of the two atomic levels  respectively,
   the  $  g $ and $  g^{\prime}= \beta g, 0 \leq \beta \leq 1 $ are the atom-photon rotating wave (RW)
   and the counter-rotating wave (CRW) coupling respectively.
   If $ \beta=0 $,  Eq.\ref{u1z2z2} reduces to the $ U(1) $ Dicke model \cite{berryphase,gold,comment} with the $ U(1) $ symmetry
   $ a \rightarrow  a  e^{ i \theta}, \sigma^{-} \rightarrow \sigma^{-} e^{ i \theta} $ leading to the conserved quantity
   $ P=  a^{\dagger} a + J_z $.  The CRW $ g^{\prime} $ term breaks the $ U(1) $ to the $ Z_2 $ symmetry
   $ a \rightarrow -a , \sigma^{-} \rightarrow -\sigma^{-} $ with the conserved parity operator  $ \Pi= e^{ i \pi ( a^{\dagger} a + J_z ) } $.
   If $ \beta=1  $, it becomes the $ Z_2 $ Dicke model \cite{chaos,qcphoton,extra}.

    After performing a rotation around the $ J_y $ axis by $ \pi/2 $, one can write $ H=H_0 + V $  where $
    H_0  =  \omega_a [ a^{\dagger} a + G ( a^{\dagger}+ a ) J_z ], G= \frac{ g ( 1 + \beta) }{ \omega_a \sqrt{N} } $ and the perturbation $
    V  =  - \frac{\omega_b}{2}[ J_+ ( 1 + \lambda ( a^{\dagger}- a )) +   J_- ( 1 - \lambda ( a^{\dagger}- a )) ] $
    where $ \lambda= \frac{ g( 1-\beta) }{ \omega_b \sqrt{N} } $ is a dimensionless parameter of order 1
    when $ 1-\beta $ is small in the large $ g $ limit.
    In principle, the strong coupling expansion is performed in the large $ g $ limit $ G \gg 1 $, but with a small $ 1-\beta $ such that
    $ \lambda $ is of order 1. In practice, as compared to ED, the method works well also when
    $ g $ is not too close to $ g_c=\frac{\sqrt{\omega_a \omega_b}}{1 + \beta } $ and
    $ \beta $ is not too close to the $ U(1) $ limit $ \beta=0 $.

   Define $  A= a + G J_z $, then $
   H_{0}  =  \omega_a [ A^{\dagger} A -( G  J_z )^{2} ] $  \cite{chaos,china}.
   Because $ [ A, J_z]=0 $, we denote the simultaneous eigenstates of $ A $ and $ J_z $ as $ | l \rangle_m | j m \rangle, m=-j, \cdots, j, l=0,1,\cdots $.
   The eigenstates satisfy $ J_z | j m \rangle = m \hbar | j m \rangle,  A^{\dagger}_m A_m | l \rangle_m= l | l \rangle_m $ where
   $ A_m= a + G m $, $   | l \rangle_{m}= D^{\dagger}(  g_{m} ) |l \rangle=D( - g_{m} ) |l \rangle $ where
   $ D(\alpha)= e^{ \alpha a-\alpha^{*} a^{\dagger} }, g_{m}=m G $ and $ |l \rangle $ is just the $ l $-photon  Fock state.
   The zeroth order eigen-energies are $  H_{0} | l \rangle_m | j m \rangle =E^{0}_{l,m} | j m \rangle,  E^{0}_{l,m}= \omega_a(l-g^{2}_m) $.
   All the eigenstates can be grouped into even or odd under the parity operator $ \Pi= e^{ i \pi ( a^{\dagger} a -J_x ) } $:
\begin{eqnarray}
  |e\rangle= \frac{1}{\sqrt{2} } [ | l \rangle_m |j, m  \rangle + (-1)^l | l \rangle_{-m} |j, -m  \rangle ]    \nonumber  \\
  |o\rangle= \frac{1}{\sqrt{2} } [ | l \rangle_m |j, m  \rangle - (-1)^l | l \rangle_{-m} |j, -m  \rangle ]
\label{eolm}
\end{eqnarray}

   The ground state is a doublet at $ | l=0 \rangle_{\pm j} |j, \pm j  \rangle $.
   In the large $ g $ limit, the excited states can be grouped into two sectors:
   (1) The atomic sector with  the eigenstates  $ | l > 0 \rangle_{\pm j} |j, \pm j  \rangle $ with the energies $ l \omega_a $.
   The first excited state $ l=1 $ with the energy $ \omega_a $ is the remanent of the pseudo-Goldstone mode in the $U(1) $ regime  \cite{gold}.
   (2) The optical sector  with  the eigenstates  $ | l \rangle_{m} |j, m  \rangle, |m | < j $.
   The first excited state has the energy $ \omega_o=E^{0}_{l,m=j-1}-E^{0}_{l,m=j} =
   \omega_a G^2( 2j-1 )= \frac{ g^2 (1+\beta)^2 }{ \omega_a } ( \frac{2j-1}{2j} ) $ and
   is the remanent of the Higgs mode in the $U(1) $ regime \cite{gold}.
   So in the strong coupling limit, there is wide separation between the atomic sector and the optical sector.
   This makes the strong coupling expansion very effective to explore the physical phenomena
   in the superradiant regime.


{\sl Ground state  ( $ l =0 $ )  splitting --- }
  The two degenerate ground state are $ | 1 \rangle= | l =0 \rangle_{-j} |j, -j  \rangle,
  | 2 \rangle=| l =0 \rangle_j |j, j  \rangle $ with the zeroth order energy $ E_0=  - \omega_a (Gj)^2 $.
  By a second order perturbation, one finds a non-zero diagonal matrix element
  $ V_{11}=V_{22}= V_0(\lambda) = - \frac{ \omega^2_b}{\omega_a} \frac{ 2 j  }{  2j-1  } \frac{ 1 + \lambda^2 }{ G^2 } < 0 $.
  However, one needs to perform a $ N = 2j $  order perturbation to find the first non-zero contribution to
  the off-diagonal matrix element  $ V_{12}=V_{21}= \Delta_0(\lambda) $:
\begin{eqnarray}
  \Delta_0(\lambda)  =   -  \frac{ N^2 \omega_b}{2} ( \frac{ \omega_b}{2 \omega_a G^2 } )^{N-1} e^{- (NG)^2/2 }
     \nonumber  \\
    \times  \sum^{N}_{l=0} \frac{ \lambda^l }{ (N-l) ! }  \sum^{[l/2]}_{n=0}
    \frac{ (-1/2)^n (-N G)^{l-2n} }{ n ! ( l-2n) ! }
\label{delta0lam}
\end{eqnarray}
  where $ [l/2] $ is the closest integer to $ l/2 $ and $
   \frac{ \lambda }{G} = \frac{ 1- \beta}{1+ \beta } \frac{\omega_a}{\omega_b} $.

\begin{widetext}

\begin{figure}
\includegraphics[width=3.5cm]{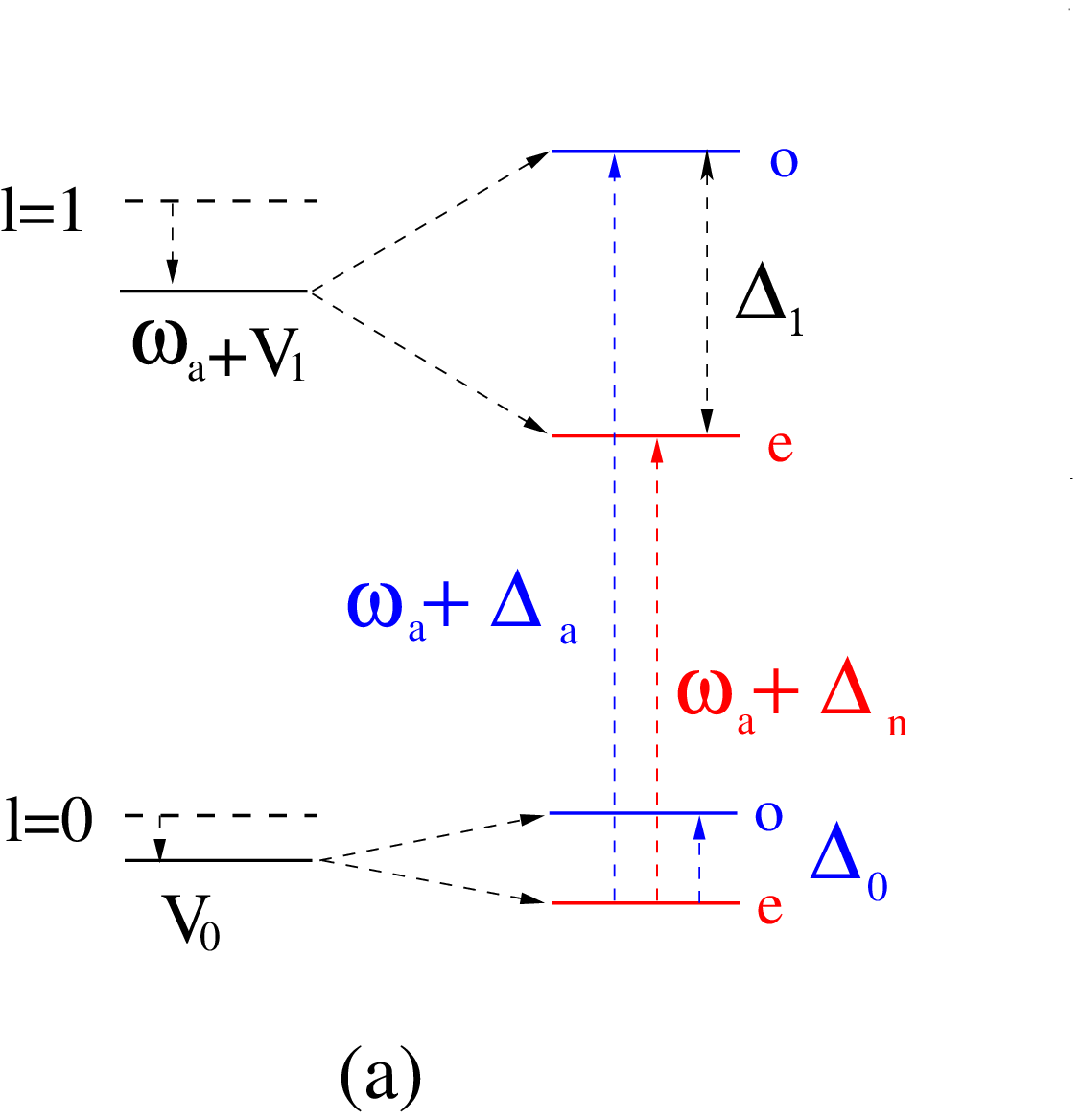}
\includegraphics[width=3.5cm]{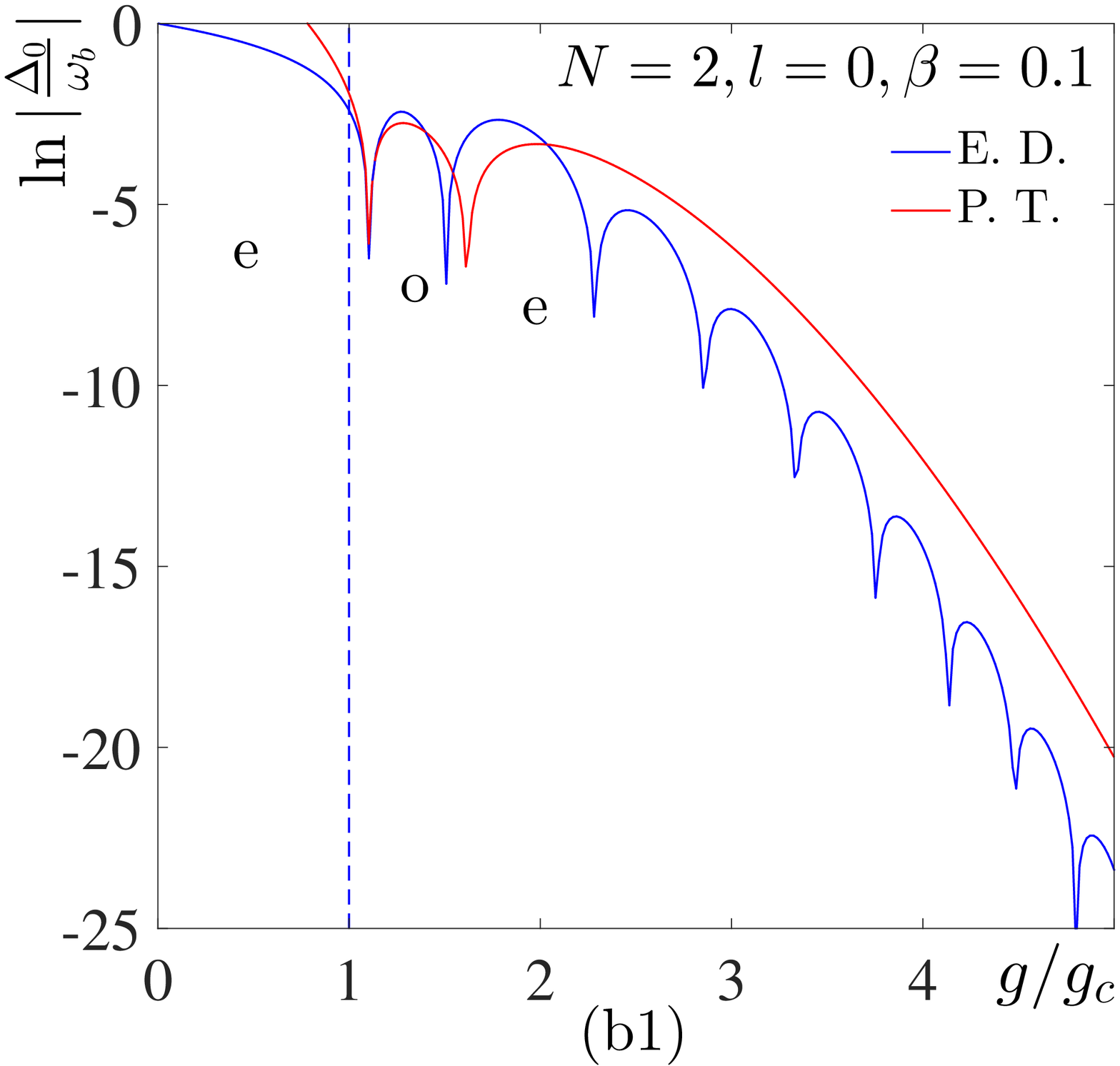}
\includegraphics[width=3.5cm]{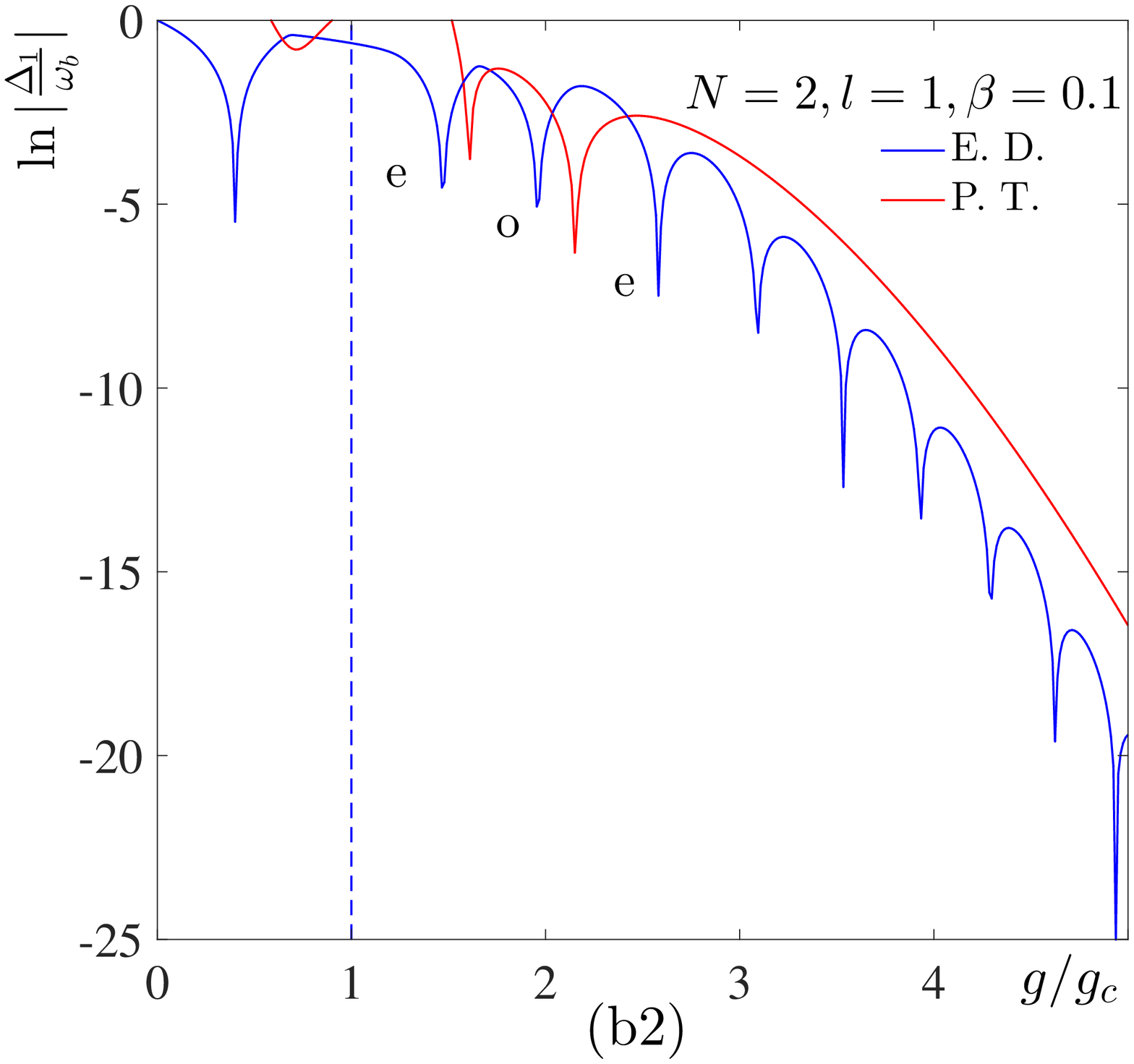}
\includegraphics[width=3.5cm]{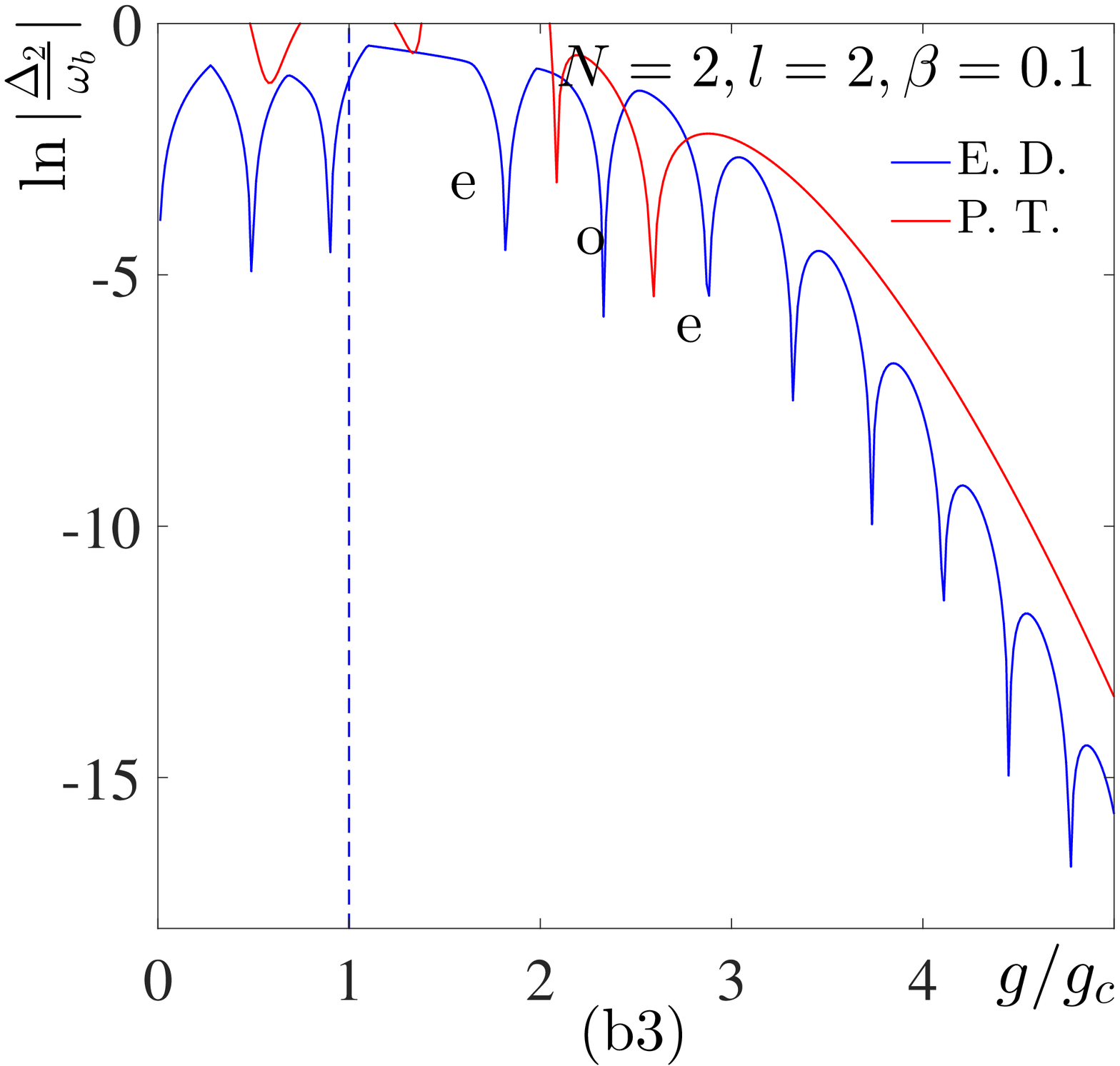}
\includegraphics[width=3.5cm]{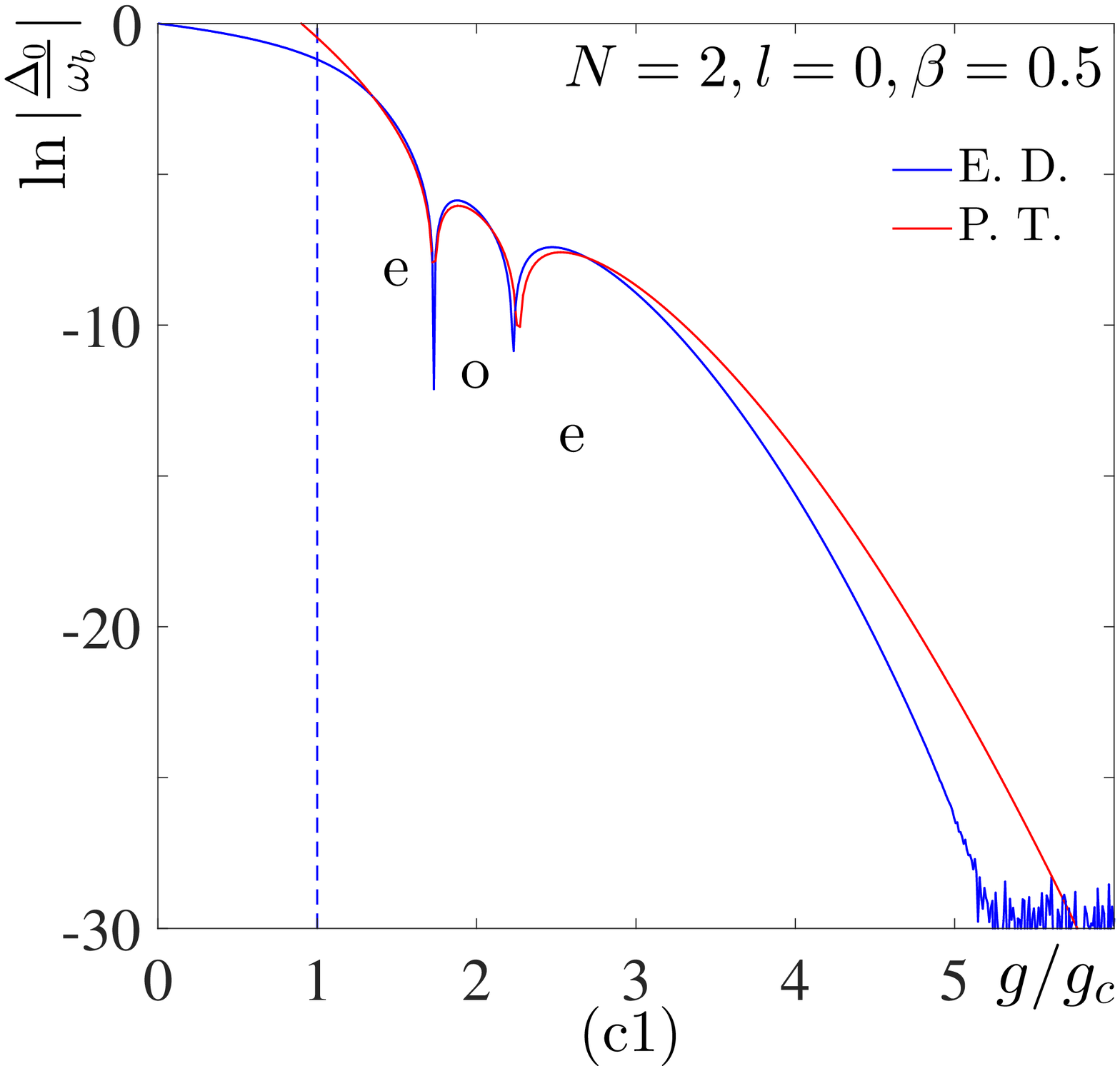}
\includegraphics[width=3.5cm]{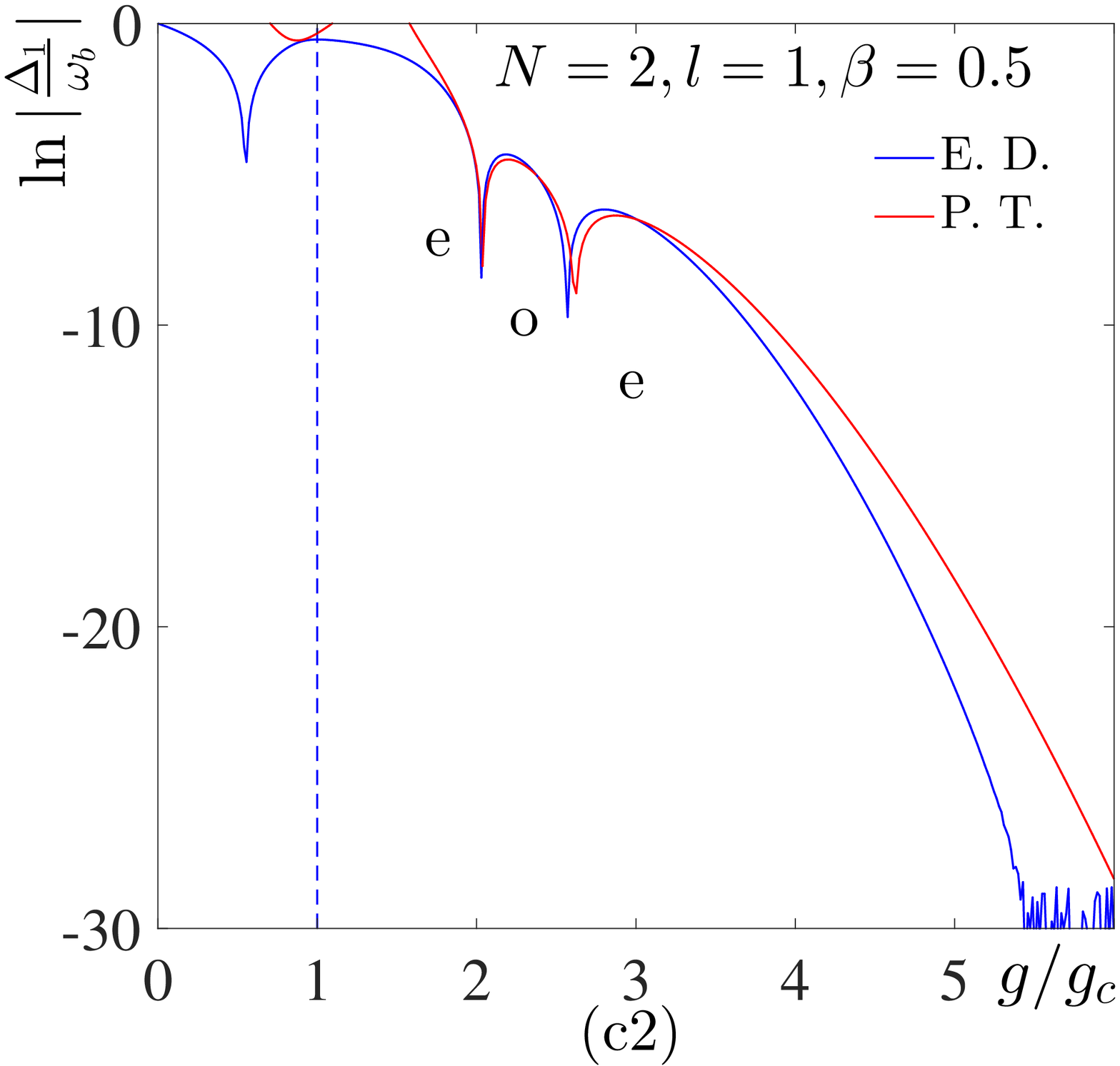}
\includegraphics[width=3.5cm]{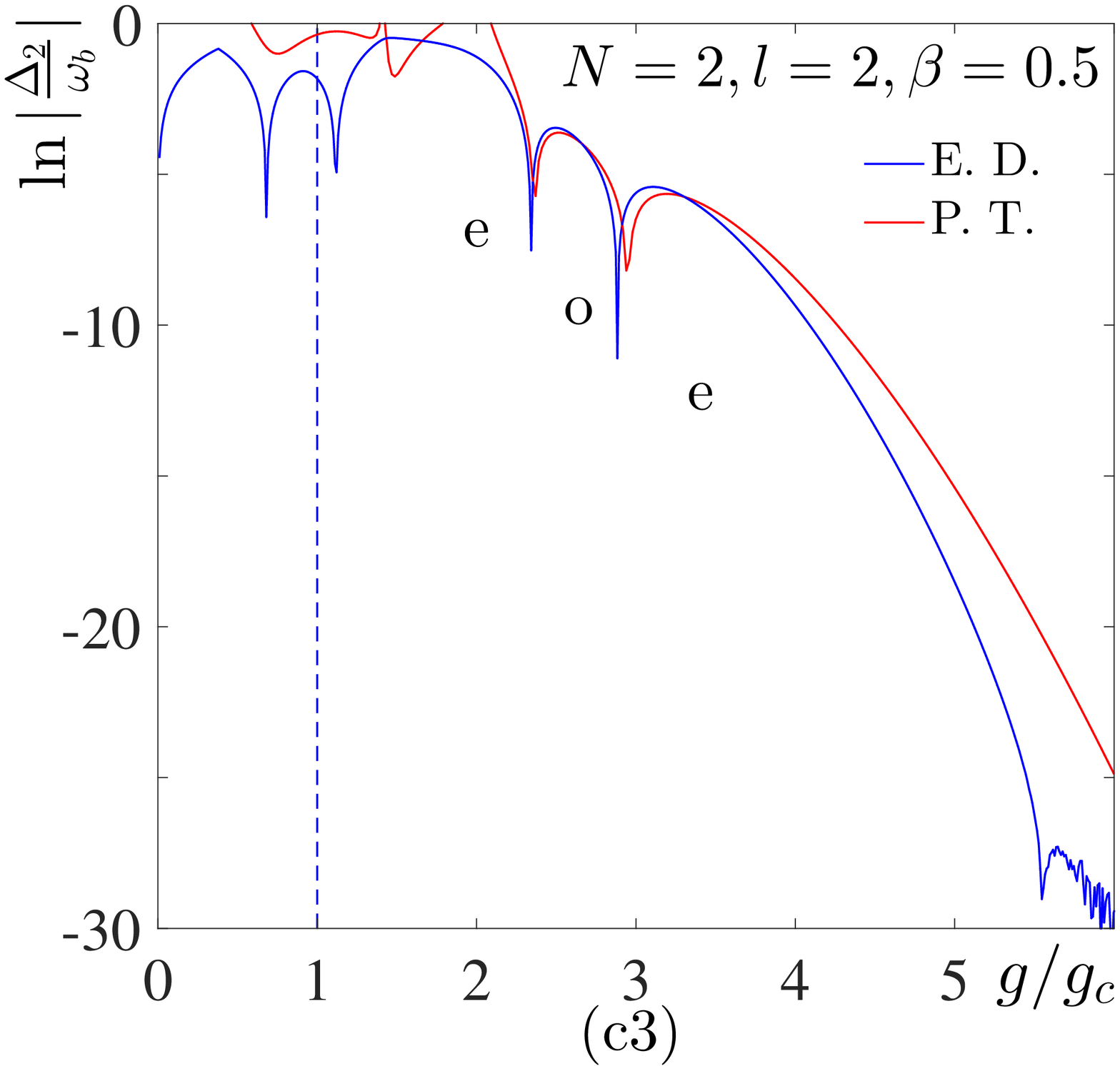}
\includegraphics[width=3.5cm]{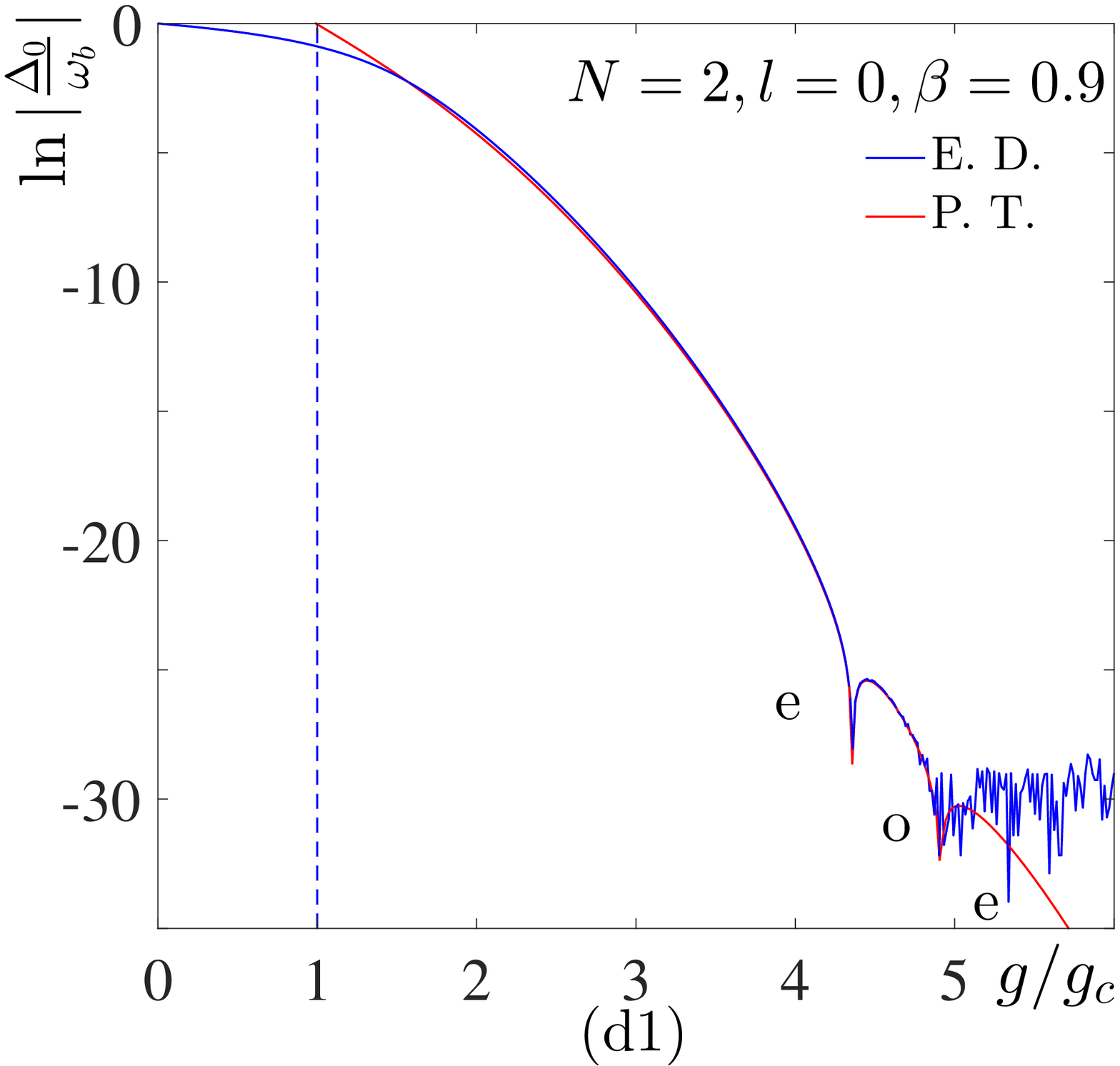}
\includegraphics[width=3.5cm]{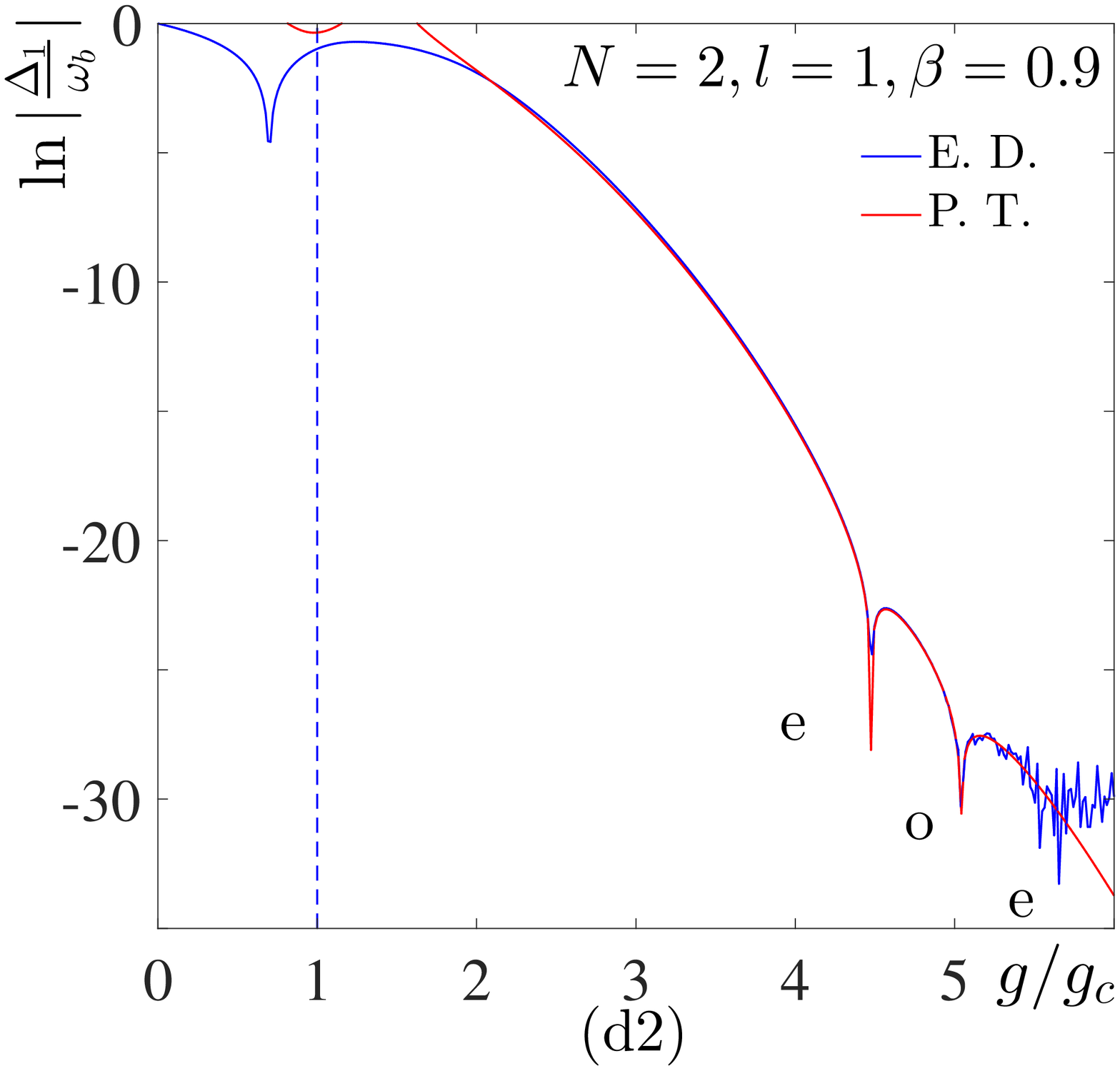}
\includegraphics[width=3.5cm]{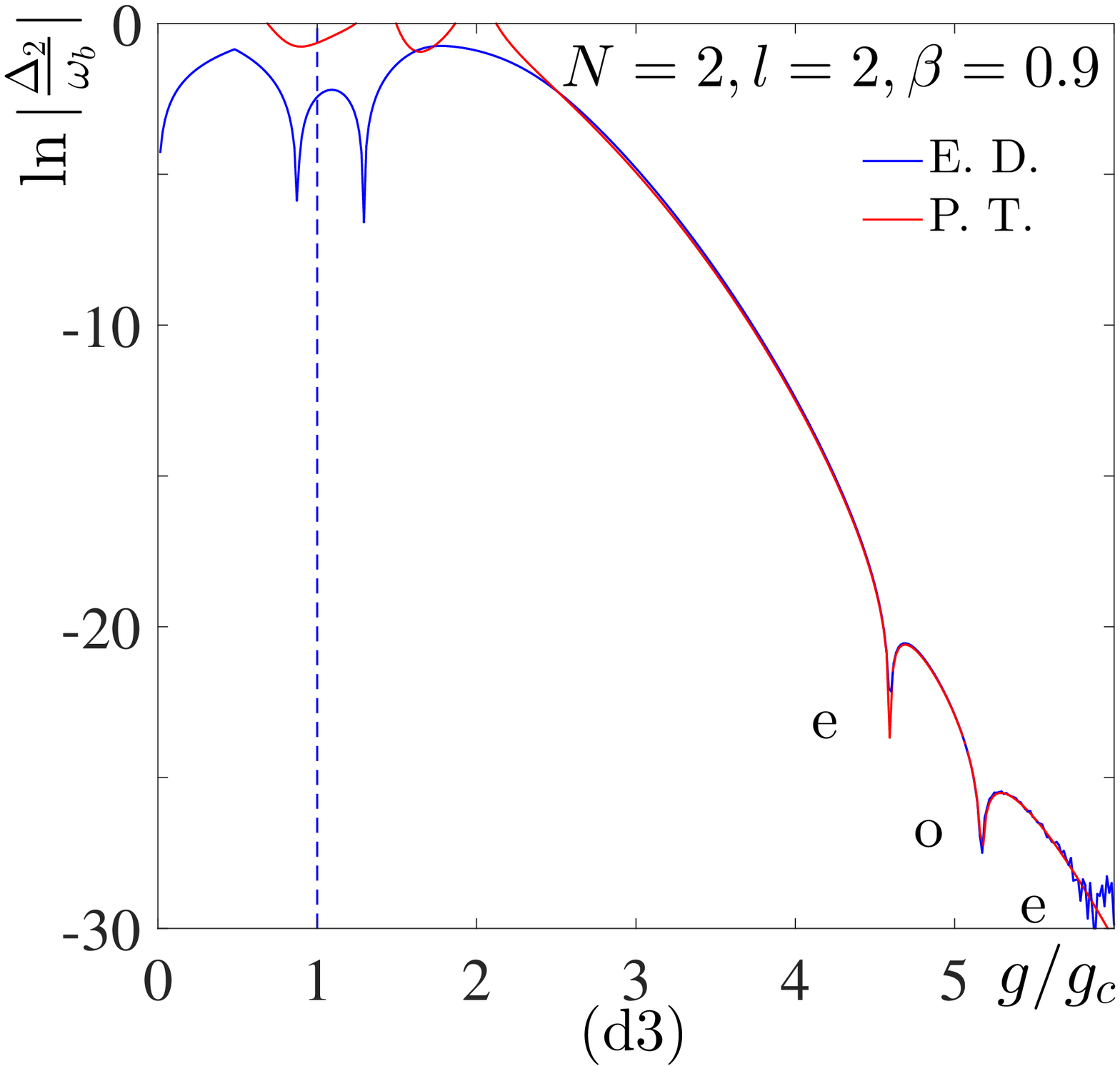}
\caption{ (Color Online) (a)  The energy shifts $ V_0<0, V_1<0 $ and splittings $ \Delta_0, \Delta_1 $ of the ground state $ l=0 $ and the first excited state $ l=1 $. Shown here is  $ \Delta_0 <0, \Delta_1 > 0 $ case where the even parity state is the ground state at $ l=0,1 $.
Being $ \Delta_0 > 0 $, the ground state  between the first two zeros in (b)-(d) has the odd parity
The blue and red transition lines can be mapped out by photon and photon number correlation functions  Eq.\ref{aaa1} and \ref{nn1} respectively.
Shown in (b)-(d) are the even/odd splitting $ \Delta_l $ for $ N=2 $ at $ \beta=0.1, 0.5,0.9£¬
l=0, 1, 2 $ in the Log scale $ \ln | \frac{\Delta_l}{ \omega_b} | $ versus $ g/g_c $.
The labels $ e $ and $ o $ are the parity of the ground states. The red ( blue ) line is from the strong coupling expansion and the ED.
The numerical sharp dips mean the zero splittings.
It always starts with the even parity with oscillating parities at $ l=0,1,2 $.
 There are also none, one and two level crossing(s) in the normal regime at $ l=0,1,2 $.
 The ED gives infinite number of zeros after the first $ N=2 $ zeros which can only
 be achieved from higher order perturbations in the strong coupling expansion.
 In (b) at $ \beta=0.1 $, the strong coupling results match well with those from the ED at $ l=0 $, but not too well at $ l=1,2 $ in the first
 $ N=2 $ zeros. Even so, they match well the envelop of the splitting
 at $ l=0,1,2 $ ( namely, the maximum splitting  ).
 In both (c) and (d), the strong coupling results match very well with those from ED in the first $ N=2 $ zeros.
 The other zeros are far apart from the first $ N=2 $ zeros and out of the scope in the figure.
 In (b) or (c), if one follow the ground state with the odd parity, there are
 some or slight shifts of zeros to the right at $ l=0,1,2 $.
 In (d), the shifts are very small as dictated by  Eq.\ref{deltalsim} in the limit $ G = \frac{g}{g_c} \frac{1}{\sqrt{N} } \gg 1 $.
 At too strong couplings, the ED may become ( noise ) un-reliable due to the cutoff introduced in the ED. }
\label{zeros159}
\end{figure}

\end{widetext}

   Setting $ \lambda=0 $ in Eq.\ref{delta0lam} leads to the splitting in the $ Z_2 $ Dicke model at $ \beta=1 $ ( Fig.S1d ):
\begin{equation}
  \Delta_0 = - \frac{ \omega_b }{ ( N-1 )! } ( \frac{ \omega_b}{ 2 \omega_a } )^{N-1} \frac{ 2 g^2 }{\omega^2_a}   e^{-N \frac{ 2 g^2 }{\omega^2_a} } <0
\label{delta0z2}
\end{equation}
   which is  always a negative quantity, so leads to the  even and odd  parity as the ground state and
   the excited state in the  $ l=0, m=j $ doublet in Eq.\ref{eolm}
   having the energies $  E_{o/e} = E_0 + V_0 \pm |\Delta_0 | $ ( Fig.\ref{zeros159}a ).

   Now we study the dramatic effects of the anisotropy $ \lambda > 0 $ encoded in Eq.\ref{delta0lam}.
   If removing the exponential factor $ e^{- (G^{\prime} )^2/2 } $ where $ G^{\prime}= N G $,
   Eq.\ref{delta0lam} is  a $ 2N $-th polynomial of $ g $. We find that it always has $ N $  positive zeros in $ g $  beyond the  $ g_{c} $
   ( namely,  fall into the super-radiant regime ).
   Higher than the $ N-$th order perturbations will lead to other zeros at larger $ g $ shown in Fig.\ref{zeros159}b.
   Any changing of sign in $ \Delta_0(\lambda) $ leads to the exchange of the parity in
   the ground state  $ l=0, m=j $ in Eq.\ref{eolm}  ( namely, Eq.S1 ) with the energies
   $  E_{o/e} = E_0 + V_0(\lambda) \pm |\Delta_0(\lambda) | $ in Fig.\ref{zeros159}a.
   So any $ \lambda > 0 $ will lead to infinite number of level crossings with alternative parities in the ground state, which is indeed observed
   in the ED results Fig.S1 for the energy levels at $ N=2 $ and $ \beta=0.1, 0.5,0.9 $.
   It is the anisotropy which leads to the parity oscillations in the superradiant regime.
   However, at $ \beta =1 $, the infinite level crossings are pushed to infinity, so no parity oscillations in Fig.S1d anymore \cite{extra}.



{\sl Doublet splitting at $ l > 0 $ ---}
   Now, we look at the energy splitting at $ l>0 $. The diagonal matric element at $ l=0 $ can be easily generalized to $ l > 0 $ case:
   $ V_{11}=V_{22}= V_l(\lambda) = - \frac{ \omega^2_b}{\omega_a} \frac{ 2 j  }{  2j-1  } \frac{ 1 + \lambda^2 ( 2l + 1 ) }{ G^2 } < 0 $.
   By performing a $ N = 2j $  order perturbation,  we also find a general ( but a little bit complicated ) expression for the
   off-diagonal matrix element $ V_{12}=V_{21}= \Delta_l(\lambda) $.
   However, in the $ G \gg 1 $ limit, it can be simplified to:
\begin{equation}
   \Delta_l(\lambda) \sim \frac{(-1)^l }{ l ! } ( G^{\prime 2} )^l \Delta_0 (\lambda)
\label{deltalsim}
\end{equation}
   where $ \Delta_0 (\lambda) $ is given in Eq.\ref{delta0lam}.
   It is enhanced due to the large prefactor $ G^{\prime 2l} $.
   Note that it is this oscillating sign $ (-1)^l $ which leads to the even/odd parity state
   with an extra $ (-1)^l $ in Eq.\ref{eolm} with $ m=j $ ( namely Eq.S2 ). The $ l$-th levels
   have the energies $  E_{o/e} = E^{0}_l + V_l(\lambda) \pm |\Delta_l(\lambda) |, E^{0}_l= \omega_a[ l- (Gj)^2 ] $ with $ l=1 $
   shown in Fig.\ref{zeros159}a.
   The diagonal part of the excited energy $ (E^{0}_l + V_l(\lambda))-(E^{0}_0 + V_0(\lambda))= l \omega_a -( |V_l|-|V_0|)= l \omega_a-
   \frac{ \omega^2_b}{\omega_a} \frac{ 2 j  }{  2j-1  } \frac{ 1 + 2 \lambda^2 }{ G^2 } <  l \omega_a $,
   but approaches $ l \omega_a $ from below in the $ G \gg 1 $ limit. This is indeed confirmed by the ED in Fig.S1.

   Eq.\ref{deltalsim} shows that at the $ N $-th order perturbation,
   the number of zeros remains to be $ N $ and the positions of the zeros are independent of $ l $ in the $ G \gg 1 $ limit.
   This observation is indeed confirmed in the following ED results in Fig.\ref{zeros159}.

{\sl Comparison with Exact Diagonziation (ED) results:---}
  In Fig.\ref{zeros159} (b)-(d), we compare Eq.\ref{delta0lam} and \ref{deltalsim} with the ED results on
  the energy level splitting between the doublets ( the Schrodinger Cats states with  even and odd parity )
  for $ N=2 $ at $\beta=0.1, 0.5,0.9, l=0,1,2 $.
  We find the first $ N $ zeros ( or parity oscillations )  from the strong coupling expansion match those from the ED
  nearly perfectly well at $ \beta=0.5,0.9 $ in the super-radiant regime.
  Of course, the ED may not be precise anymore when $ g $ gets too close to the upper cutoff introduced in the ED calculation
  as shown in Fig.\ref{zeros159}d. In fact, the first $ N=2 $ zeros of  Eq.\ref{delta0lam} can be found exactly
  as the two positive roots $ G_{\mp}= ( \sqrt{ 1 + 8 \frac{ 1+\beta }{1-\beta } } \mp 1 )/4 $ falling in the superradiant regime.
  The spacing between the two roots $ \Delta( \frac{g}{g_c} )=\frac{1}{\sqrt{2}} $ is independent of $ \beta $ as shown in Fig.\ref{zeros159}c,d.  As $ \beta \rightarrow 1^{-} $, both roots $ \sim (1-\beta)^{-1/2} $ are pushed into the infinity.

   Eq.\ref{deltalsim} is also confirmed by the ED shown in Fig.\ref{zeros159}d
   for $ N=2, \beta=0.9, l=0,1,2 $ where the positions of the first $ N=2 $ zeros only depend on $ l $ very weakly.
   So between the two zeros, at $ l=0,1,2,\cdots $, the energy levels are in the pattern
   $ (e,o),(e,o),\cdots $  when $ \Delta_0(\lambda) <0 $ shown in Fig.\ref{zeros159}a  ( or $ (o,e),(o,e),\cdots $ when $ \Delta_0(\lambda) > 0 $ ).

{\sl Photon, squeezing and number correlation functions --- }
    In order to calculate the photon correlation functions in the strong coupling limit, one not only needs to find the energy
    levels as done in the previous sections and in Fig.\ref{zeros159}a, but also the wavefunctions listed in the SM.
  Using the Lehmann representations,  we find there is no first order correction to the  normal photon correlation function, but
  there is one $ \sim 1/G^2 $ to the anomalous  photon correlation function:
\begin{eqnarray}
  \langle a( t ) a^{\dagger}(0) \rangle &  =  & A e^{- i|\Delta_0| t} +  e^{- i( \omega_a + \Delta_a )t }
        \nonumber  \\
  \langle a( t ) a(0) \rangle & = &  A e^{-i |\Delta_0|t}
  -B e^{- i (\omega_a + \Delta_a ) t }
\label{aaa1}
\end{eqnarray}
  where $ A= ( G j )^2= N \frac{ g^2 (1+\beta)^2 }{ 4 \omega^2_a } \sim G^2 $ is the photon number in the ground state \cite{photonnumber} and
  $ B= \frac{ \lambda^2 }{ G^{2} } (\frac{ \omega_b}{2\omega_a})^2 \frac{ 2j}{2j-1 } \sim 1/G^2  $
  and $ \Delta_a=(V_1-V_0)+ \frac{1}{2}( |\Delta_1| + |\Delta_0| ) $ shown in Fig.\ref{zeros159}a .
  One can see the anomalous spectral weight $ - B \sim (\lambda/G)^2 $ is negative and completely due to $ \lambda $  ( away from the $ Z_2 $ limit ).
  So the $ B $ term in the anomalous  photon correlation function can reflect precisely the anisotropy $ \beta $ and
  can be easily detected in phase sensitive Homodyne measurements \cite{exciton}.

   Similarly, we also find the first order correction $ \sim 1/G^2 $ to the photon number correlation function:
\begin{equation}
  \langle n( t ) n(0) \rangle -  \langle n \rangle^2 =
  A [1 + B ]^2  e^{- i( \omega_a + \Delta_n )t }
\label{nn1}
\end{equation}
  where $ \Delta_n=(V_1-V_0)- \frac{1}{2}( |\Delta_1| - |\Delta_0| ) $ shown in Fig.\ref{zeros159}a and
  $ \langle n \rangle = A $ is the photon number in the ground state which does not receive first-order correction.
   From Eq.\ref{aaa1} and \ref{nn1}, one can see that the $ \Delta_0 $ can be directly extracted from the very first
   frequency in Eq.\ref{aaa1}, while $ |\Delta_1|= \Delta_a - \Delta_n $ and $ V_1-V_0= ( \Delta_a+ \Delta_n )/2 - |\Delta_0| $.
   So all the parameters of the cavity systems such as the doublet splittings $ \Delta_0( \lambda), \Delta_1( \lambda) $ and
     energy level shifts $ V_1-V_0 $ in Fig.\ref{zeros159}a
     can be extracted from the photon normal and anomalous Green functions Eq.\ref{aaa1}  and photon number correlation functions
     Eq.\ref{nn1}. They can be measured by photoluminescence,
     phase sensitive homodyne and  Hanbury-Brown-Twiss ( HBT ) type of experiments \cite{exciton} respectively.



{\sl Experimental realizations: }
    There have been extensive efforts to realize the Schrodinger Cat state
  in trapped ions \cite{cat1} and superconducting qubit systems \cite{cat2}.
  Here, the Schrodinger Cats in Eq.\ref{eolm} with $ l=0,1,2...$ and $ m=j $ can be
  prepared in the superradiant regime, its size can be continuously tuned from $ N \sim 3-9 $,
  it involves all the $ N $  number of atoms ( qubits ) and photons strongly coupled inside the
  cavity and could have
  important applications in quantum information processions.

   In order to observe the parity oscillation effects, one has to move away from the $ Z_2 $ limit
   realized in the experiments \cite{orbitalt,orbital,switch}, namely, $ 0 < \beta < 1 $. This has been realized in the recent experiment
   \cite{expggprime} with cold atoms inside an optical cavity which can tune $ \beta $ from $ 0 $ to $ 1 $.
   It should be straightforward to reduce the number of atoms to a few \cite{fewboson,fewfermion}.
   In circuit QED systems, there are various experimental set-ups such as charge,
   flux, phase qubits or qutrits, the couplings could be capacitive or inductive through $ \Lambda, V, \Xi $ or the $ \Delta $ shape \cite{you}.
   Especially, continuously changing $ 0 < \beta < 1 $ has been achieved in the recent experiment \cite{qubitstrong}.
   An shown in \cite{gold}, the repulsive qubit-qubit interaction also reduces the critical coupling  $g_c$.
   
   From Fig.\ref{zeros159}b1, at $ N=2, \beta=0.1, l=0 $, one can estimate the maximum splitting between
   the first two zeros $ \Delta_0 \sim 0.1 \omega_a $ which is easily experimentally measurable. $ \Delta_l $ increases as $ l=1,2 $ as shown in Fig.\ref{zeros159}b2,b3.
   At $ \beta=0.5 $ in Fig.1c1, $ \Delta_0 $ decreases to $ \sim 0.01 \omega_a $ which is still easily measurable.
   At $ \beta=0.9 $ in Fig.\ref{zeros159}d, $ \Delta_0 $ decreases to $ \sim 10^{-11} \omega_a $ which may become difficult to measure.
   However, in view of recent advances in the precision measurements in the detection of the elusive gravitational waves \cite{ligo},
   it is also possible to measure
   such a tiny splitting by the phase sensitive homodyne detection \cite{switch}. 
   So the parity oscillations can be easily experimentally measured when $ \beta $ is not too close to the $ Z_2 $ limit.

{\sl Conclusions and discussions: }
  The four standard quantum optics models at a finite $ N $ were proposed by the old generation of great physicists  many decades ago.
  Their importance in quantum and non-linear optics
  ranks the same as the bosonic or fermionic Hubbard models and
  Heisenberg models in strongly correlated electron systems and the Ising models in Statistical mechanics \cite{aue}.
  Despite their relative simple forms and many previous theoretical works,
  their solutions at a finite $ N $, especially inside the superradiant regime, remain unknown.
  In this work, we addressed this outstanding historical problem by using the strong coupling expansion and ED.
  We are able to analytically calculate various photon correlation functions in the superradiant regime remarkably accurate
  except when $ \beta $ is too small where the (non)-degenerate perturbations
  near the $ U(1) $ limit ( $ \beta=0 $ ) works well \cite{gold}.
   The present work may inspire several new directions.
   From the wavefunctions listed in SM, it would be interesting to evaluate \cite{un} the effects of the parity oscillations on
   the atom-photon entanglements in the Schrodinger cats at a given $ l=0,1,2...$.
   It is important to incorporate the effects of the external pumping and cavity photon decays \cite{exciton} to study
   the de-coherences of the Schrodinger cats in the non-equilibrium $ U(1)/Z_2 $ Dicke model.
   It would be tempting to study the arrays of cavities leading to the
   $ Z_2/U(1) $ Dicke lattice models \cite{hakan} with general $ 0 < \beta < 1 $.
 We acknowledge  NSF-DMR-1161497, NSFC-11174210 for supports. The work at KITP was supported by NSF PHY11-25915.
CLZ's work has been supported by National Keystone Basic Research Program (973 Program) under Grant No. 2007CB310408,
No. 2006CB302901 and by the Funding Project for Academic
Human Resources Development in Institutions of Higher Learning Under the Jurisdiction of Beijing Municipality.



\begin{thebibliography}{99}
\bibitem{extra}  It is tempting to try to sort out if there is an extra symmetry at the $ Z_2 $ limit $ \beta=1 $.
Taking the Eq.\ref{u1z2z2}, one can see the last two terms exchange under the transformation $ a \rightarrow e^{i \pi/2} a,
\sigma_{-} \rightarrow e^{i \pi/2} \sigma_{-} $.
It is also easy to see the Hamiltonian has the symmetry $ {\cal T} \times R(y,\pi) $ or
$ {\cal T} \times R(x,\pi) \times Z^{a}_2 $ where  $ {\cal T}  $ is the Time reversal symmetry,
$ R(y,\pi), R(x,\pi) $ are the spin rotations by $ \pi $ around $ y $ or $ x $ axis respectively and $ Z^{a}_2 $  is $ a \rightarrow -a $.
So there is no extra symmetry at the $ Z_2 $ limit.
\bibitem{photonnumber} In the first line in Eq.\ref{aaa1}, setting $ t=0 $ and using $ [ a, a^{\dagger} ]=1 $, one can also see the
photon number in the ground state $ \langle n \rangle = A $.
\end{thebibliography}
\end{document}